\documentclass[12pt,3p]{elsarticle}
\usepackage{amsmath,amsfonts,amsthm,amssymb,indentfirst,epic,url,graphicx,needspace,multirow}
\pdfoutput=1
\usepackage{moreverb,url}
\usepackage{subfigure, float}
\usepackage{multicol}
\usepackage{threeparttable}

\usepackage[colorlinks,bookmarksopen,bookmarksnumbered,citecolor=red,urlcolor=red]{hyperref}

\newcommand{\bzero}{\textbf{0}}

\newcommand{\bx}{\textbf{x}}

\newcommand{\bbeta}{ \mbox{\boldmath $ \beta $} }

\newcommand{\bSigma}{ \mbox{\boldmath $\Sigma$} }

\newcommand{\N}{\mathcal{N}}
\newcommand{\apoe}{APOE$\varepsilon$4}

\begin{document}
\begin{frontmatter}


\title{The relative efficiency of time-to-progression and continuous measures of cognition in pre-symptomatic Alzheimer's}

\author[atri]{Dan Li}
\author[atri,ug]{Samuel Iddi}
\author[atri]{Paul S. Aisen}
\author[ucsd]{Wesley K. Thompson}
\author[atri]{Michael C. Donohue\corref{mycorrespondingauthor}}
\cortext[mycorrespondingauthor]{Corresponding author}
\ead{mdonohue@usc.edu}
\author{for the Alzheimer's Disease Neuroimaging Initiative\fnref{adni}}

\address[atri]{Alzheimer's Therapeutic Research Institute, University of Southern California, San Diego}
\address[ug]{University of Ghana, Accra}
\address[ucsd]{University of California, San Diego}

\fntext[adni]{Data used in preparation of this article were obtained from the Alzheimer's Disease
Neuroimaging Initiative (ADNI) database (\url{adni.loni.usc.edu}). As such, the investigators
within the ADNI contributed to the design and implementation of ADNI and/or provided data
but did not participate in analysis or writing of this report. A complete listing of ADNI
investigators can be found at: \url{http://adni.loni.usc.edu/wp-content/uploads/how_to_apply/ADNI_Acknowledgement_List.pdf}}

\begin{abstract}
Pre-symptomatic (or Preclinical) Alzheimer's Disease is defined by biomarker evidence of fibrillar amyloid beta pathology in the absence of clinical symptoms. Clinical trials in this early phase of disease are challenging due to the slow rate of disease progression as measured by periodic cognitive performance tests or by transition to a diagnosis of Mild Cognitive Impairment. In a multisite study, experts provide diagnoses by central chart review without the benefit of in-person assessment. We use a simulation study to demonstrate that models of repeated cognitive assessments detect treatment effects more efficiently compared to models of time-to-progression to an endpoint such as change in diagnosis. Multivariate continuous data are simulated from a Bayesian joint mixed effects model fit to data from the Alzheimer's Disease Neuroimaging Initiative. Simulated progression events are algorithmically derived from the continuous assessments using a random forest model fit to the same data. We find that power is approximately doubled with models of repeated continuous outcomes compared to the time-to-progression analysis. The simulations also demonstrate that a plausible informative missing data pattern can induce a bias which inflates treatment effects, yet 5\% Type I error is maintained.
\end{abstract}

\begin{keyword}
clinical trial simulations \sep Alzheimer's Disease \sep Cox proportional hazards model \sep longitudinal data \sep mixed model of repeated measures (MMRM)\sep statistical power \sep common close design \sep Bayesian joint mixed effect model
\end{keyword}

\end{frontmatter}

\section{Introduction}
Pre-symptomatic (or Preclinical) Alzheimer's Disease (PAD) is defined by evidence of abnormal levels of fibrillar amyloid beta in brain as measured by positron emission tomography (PET) brain scan or cerebrospinal fluid (CSF) assay \cite{sperling2011toward}. Clinical trials have been initiated in this early phase of disease with the hope that, as in other diseases, early interventions will be more successful in slowing progression \cite{sperling2014a4, EARLY, caputo2017rationale}. 

In PAD, progression is typically measured by continuous assessments such as the Preclinical Alzheimer's Cognitive Composite (PACC), a cognitive performance assessment sensitive to amyloid-related decline \cite{donohue2014preclinical}. An alternative measure of progression is transition from normal cognition to Mild Cognitive Impairment (MCI). The diagnosis of MCI is not algorithmic. It is based on an expert clinician's subjective impression of clinical tests and interviews with participants or study partners. In contrast to cancer progression or death, the cognitive diagnosis (normal or MCI) can vary from one clinician to the next, or from one study visit to the next.  In a multicenter study, the diagnosis made by a clinician at a trial performance site may be confirmed by experts centrally based on review of assessments without the benefit of direct in-person assessment.

Some researchers prefer the inherent clinical meaningfulness of time-to-MCI analysis. Undoubtedly, for a given subject, a transition from normal cognition to MCI is more clinically meaningful than a point change in a continuous cognitive performance measure. However, in a clinical trial, we are still left to determine how large a randomized group difference in the rate of, or delay in, a clinically meaningful event is itself clinically meaningful.

The typical Alzheimer's clinical trial assesses cognition at clinic visits conducted every three or six months. With a continuous outcome, the primary contrast is estimated at the last scheduled visit, at 4.5 years say. Proponents of time-to-progression argue that the endpoint allows for a \emph{common close design}, similar to oncology studies, in which follow-up can continue until the last subject enrolled reaches the 4.5 year visit. The Cox Proportional Hazards model \cite{cox1972regression} admits data collected under such a design. Linear mixed-effects models can also admit data from a common close design, but assumptions about the mean trend (e.g. quadratic time trends) are necessary, similar to the proportional hazards assumption.

Some related work has demonstrated the advantages of analyzing continuous outcomes, when available, over time-to-event outcomes in other contexts. Donohue, et al. \cite{donohue2011relative} reviewed the literature and provided an analytic demonstration that, under general conditions, a mixed-effect model comparison of rate of change on a continuous outcome is effectively always more powerful than an analysis of time-to-threshold. The authors also conducted simulations based on Alzheimer's Disease Neuroimaging Initiative (ADNI) MCI subjects and demonstrated that the marginal linear model and linear mixed models are more robust and efficient than the Cox model of time from MCI to dementia. 

Our goal is to extend our earlier  work in the MCI population \citep{donohue2011relative} to the earlier biomarker defined PAD population. Specifically, we aim to compare the performance of models of repeated measures of the PACC versus time-to-progression when evaluating treatment effects in randomized trials, and to assess bias due to informative missingness. We also compare the common close design and the fixed follow-up design. We apply the Mixed Models of Repeated Measures (MMRM) \cite{mallinckrodt2003assessing} for the analysis of change in the PACC score. Constrained longitudinal data analysis (cLDA) models \cite{LiangandZeger(2000)} are also used to model the PACC scores treating time as a continuous variable. Cox proportional hazards model is applied to time-to-event endpoint.



\section{Data}
\label{data}
ADNI is a prospective observational cohort study, led by Principal Investigator Michael W. Weiner, MD, which is tracking cognitive, imaging, and biofluid markers of Alzheimer's in volunteers diagnosed as cognitively normal (CN), subjective memory concern (SMC), mild cognitive impairment (MCI) and mild-to-moderate dementia. To simulate both longitudinal continuous markers and time-to-MCI for a Preclinical AD (PAD) clinical trial, we first model the disease markers and clinical diagnosis using data from PAD ADNI participants. The PAD population is defined by a diagnosis of CN or SMC at baseline, and florbetapir PET standardized uptake value ratio (SUVR) above 1.11 \citep{Landauetal(2012)} or CSF amyloid beta (A$\beta$) below 950.6 pg/ml. The CSF threshold of 950.6 pg/ml was selected because it yields the same proportion of PAD as the 1.11 SUVR threshold. Follow-up observations, including a site clinician's diagnosis of CN, MCI, or dementia, are collected every three, six, or 12 months. For more information on the study design of ADNI, including protocols, see \url{adni.loni.usc.edu}.

Sensitive tests of cognition may show changes in PAD many years before the onset of functional decline \citep{donohue2014preclinical, donohue2017association}. In this work, we focus on seven cognitive outcomes in the PAD population, namely:
\begin{enumerate}
\item ADAS Delayed Word Recall (ADAS-DWR) \cite{mohs1988alzheimer},
\item Logical Memory Paragraph Recall (LogMem) \cite{wechsler1987manual}, 
\item Trail Making Test Part B (Trails B) \cite{tombaugh2004trail}, 
\item Mini-Mental State Examination (MMSE) \cite{folstein1975mini}, 
\item Category Fluency - Animals, 
\item Clinical Dementia Rating - Sum of Boxes (CDRSB) \cite{morris1993clinical} and 
\item Functional Assessment Questionnaire (FAQ). 
\end{enumerate}
Baseline covariates considered include age and carriage of an apolipoportein E4 (\apoe) allele. The PAD population includes a total of $N=$163 individuals, in which $N=$39 (23.9\%) were observed to progress to MCI over a median follow-up time of 4.0 years (interquartile range 2.1 to 5.6 years; maximum 11.5 years). Baseline characteristics of the modeled PAD cohort are presented in Table \ref{summaries}.

\section{Methods}
\label{methods}

\subsection{Joint mixed-effects model for longitudinal data}

To a derive a model to simulate plausible data, we first fit a model to observed ADNI data. We apply a joint (or multivariate) mixed-effects model (JMM) to simultaneous model continuous longitudinal data for disease markers in the PAD population. The model respects the within-subject correlation over time and among the battery outcomes.

Suppose we have a set of $n$ subjects followed over a time interval $[0, \tau)$. The $i$th subject provides a set of longitudinal quantitative measurements $\{$$y_{ijk}$, $j = 1,\cdots,n_{ik}$, $k=1,\cdots,p$$\}$ at time points $\{$$t_{ijk}$, $j = 1,\cdots, n_i$, $k=1,\cdots,p$$\}$.
Linear mixed-effects models are commonly used to model continuous longitudinal data. The multivariate mixed-effects model is specified as $y_{ijk}=\bx_{ijk}^{\prime}\bbeta_k + b_{0ik}+b_{1ik}t_{ijk} + \varepsilon_{ijk}$, where $\bbeta_k$ are fixed-effect regression coefficients, $b_{0ik}$ and $b_{1ik}$ are the subject- and outcome-specific random intercept and slope for individual $i$ and outcome $k$. The random effects are assumed to follow a multivariate Gaussian distribution with mean vector $\bzero$ and variance-covariance matrix $\bSigma$ with dimension $2p$, that is $\left(b_{0i1},\cdots,b_{0ip},b_{1i1},\cdots,b_{1ip} \right)^{\prime} \sim \N\left(\bzero,\bSigma\right)$. The model with multivariate random effects has the advantage of reflecting the dependency within subjects and among outcomes. The $\varepsilon_{ijk}\sim\N\left(0, \sigma^2_k\right)$ is a measurement error term, which accounts for outcome-specific variance.

Since the outcomes are in different scales, we transform the raw outcome measures into a quantile scale ranging from 0 to 1 (least impaired to most severe dementia). Quantiles are calculated using the empirical cumulative distribution function using weights that are inversely proportional to the number of observations from each diagnostic category for each outcome. The quantiles were then transformed by the inverse Gaussian quantile function resulting an approximate $Z$-score before submitting to the model. When simulating data from these models, the simulated $Z$-scores can then be back transformed to the original scale, which is integer valued for some outcomes.

Bayesian estimation is performed via Markov Chain Monte Carlo (MCMC) sampling using the \texttt{stan\_mvmer} function in R package \texttt{Rstanarm} \cite{rstanarm}. Because the \texttt{stan\_mvmer} function is limited to a maximum of three outcomes, we have coded our own version allowing up to twenty outcomes (available from \url{github.com/mcdonohue/rstanarm}).

\subsection{Random forest algorithm for diagnosis of MCI}
In order to simulate a clinician's diagnosis of MCI or worse impairment, we first use ADNI data to learn an algorithm to approximate this decision. The random forest algorithm \cite{Breiman(2001)} is an ensemble learning method for classification and regression. It operates by generating several decision trees and aggregating them. It provides reasonable and easily interpretable model when a large number of predictors are present in the data and enables applications with mixed data-types such as continuous and categorical data.

In our application, clinician diagnosis of normal cognition versus MCI or worse impairment is the binary outcome variable, and the seven continuous markers, age and education are the predictors. The model is fit using the R package \texttt{randomForest} \cite{randomForest}. The fitted model is then applied to simulated continuous outcomes to predict a clinician's diagnosis.

\subsection{Competing clinical trial models for continuous and time-to-event outcomes in simulation study}
\label{competing_models}
The simulated treatment effect on time-to-progression is modeled by the Cox proportional hazards model. For the continuous PACC, we consider MMRM and the the constrained longitudinal data analysis (cLDA) proposed by Liang and Zeger \cite{LiangandZeger(2000)}. Like most likelihood-based approaches for longitudinal data, all three models assume any missing data are missing at random (MAR).

The PACC is used as the continuous outcome measure for the PAD trials simulation study. The version of the PACC used in the study is a composite of four assessments: ADAS-DWR, LogMem, log transformation of Trails B, and MMSE. Each of the four component scores is first centered by subtracting the baseline sample mean and then divided by the baseline sample standard deviation of that component, to form standardized $Z$-scores. These $Z$-scores are averaged to form the composite. 

The MMRM for treats change from baseline in the PACC score as the outcome and baseline PACC as a predictor. It treats time as a categorical variable, which allows general mean trends in each group. MMRM has been extensively used for testing treatment effects at specific time points in clinical trials, since participants are often evaluated at a fixed and relatively small number of time points \cite{Siddiqui_etal(2009)}. In our simulation study, the within-subject dependence is modeled by a first-order autoregressive covariance structure. 

We also explore models that treat time as a continuous variable. In cLDA, the baseline outcome is treated as a response variable rather than a covariate, and constrained to have equal mean at baseline across treatment groups \cite{Liu_etal(2009), Lu(2010)}. We explore models with linear or quadratic time trends for each group.

\subsection{Simulation set-up}
\label{setup}
We conduct a simulation study to evaluate the performance of the competing models described in Section \ref{competing_models}. In each of 1000 simulated clinical trials with visits every 6 months from 0 to 8 years, a total of 1000 and 1500 patients are respectively randomized to either treatment or placebo in 1:1 ratio. We also assume the proportion of MCI progressors is 24\% (based on ADNI data, as noted above). 

For the placebo group, no changes will be made to the JMM fit to ADNI. For the treatment group, we will impose large (40\% improvement on rate of change over the control), moderate (30\% improvement), small (20\% improvement) and null (same as the control) treatment effects on all outcomes. The PACC scores are calculated by taking the average of the four simulated component $Z$-scores.

To simulate non-ignorable missing data, three dropout categories are considered: intolerability, inefficacy and missing completed at random (MCAR). Participants having intolerability or inefficacy drop out from the study immediately after six and twelve months, respectively. For MCAR, we assume linear attrition rate of 5\% per year for both the treatment and placebo groups. The simulated dropout rates are:
\begin{itemize}[noitemsep,leftmargin=*]
    \item[-] Treatment group:
	\begin{itemize}[noitemsep,leftmargin=*]
		\item[-] Null: inefficacy (15\%), intolerability (10\%), MCAR (5\%/year attrition rate);
		\item[-] Alternative: inefficacy (8\%), intolerability (10\%), MCAR (5\%/year attrition rate);
	\end{itemize}
	\item[-] Placebo group: inefficacy (15\%), MCAR (5\%/year attrition rate).
\end{itemize}
In order to assess bias due to missing data, we simulate complete data for every subject. The complete data is appropriately censored for the analysis of ``observed'' data, and left uncensored for analysis of the ``complete'' data. Completers and MCAR dropouts are assumed to have the same longitudinal mean profile within each treatment arm. Dropouts due to intolerability are simulated to have the expected benefit, on average, until dropout, followed by an ``unobserved'' benefit that is diminished by a factor of 15\%. Dropouts due to inefficacy are simulated to have no benefit.

The four competing clinical trial models are MMRM, $\textrm{cLDA}^1$ (linear) and $\textrm{cLDA}^2$ (quadratic) for continuous PACC scores; and Cox for time-to-progression, with two baseline covariates: age at baseline and carriage of the \apoe allele. The Cox model will use all data observed out to 8 years until the last subject reaches the final scheduled visit under the common close design. We assume a linear enrollment rate such that enrollment is completed in 4 years and about half the subjects contribute ``extra'' common close follow-up in the 4.5 to 8 year range to the Cox model. The MMRM, $\textrm{cLDA}^1$ and $\textrm{cLDA}^2$ will only use data up to last scheduled visit, i.e., from 0 to 4.5 years. 

We focus on ``treatment policy'' estimands of interest. The estimand will be the difference between randomized groups in the intention-to-treat population in terms of either: (I) Rate (hazard ratio) of progression to MCI/Dementia (Cox); (II)  Group difference in PACC at final study time point (MMRM and $\textrm{cLDA}^1$); or (III) Area between mean PACC curves ($\textrm{cLDA}^2$). We show how to carry out the hypothesis test of case (III) in the Appendix. Let $Y_{ijk}$ denote the simulated PACC scores for subject $i$ randomized to group $j$ at time point $k$, where $i=1,\cdots,n_j$, $j=D,P$ and $k=1,\cdots,T$. And $k=0$ represents the baseline time point, $D$ is the treatment group and $P$ is the placebo group. If the estimand of interest is the change from baseline at time $T$, i.e., $Y_{ijT}-Y_{ij0}$. The object is to estimate the between-treatment difference $\delta=\mu_{P}-\mu_{D}$, where $\mu_j=E\left(Y_{ijT}-Y_{ij0}\right)$. A two-tailed test $H_0:\delta=0$ versus $H_1:\delta\neq 0$ is carried out to evaluate whether treatment is different from placebo.

For each simulated dataset, we apply all four competing models to calculate point estimates of $\delta$ using the observed data (i.e., $\delta_{\textrm{obs}}$) and the complete data (i.e., $\delta_{\textrm{comp}}$). For each model, ``bias'' is calculated as the median of the 1000 point estimates of $\delta_{\textrm{obs}}$ minus $\delta_{\textrm{comp}}$; ``bias in percent'' is computed as the median of the 1000 points estimates of  $\delta_{\textrm{obs}}$ minus $\delta_{\textrm{comp}}$ and then divided by $\delta_{\textrm{comp}}$. The interquartiles $\mathrm{Q}_1$ and $\mathrm{Q}_3$ are also summarized.

In a real clinical trial, the endpoint is measured for completers but is missing for those who either drop out from the study either due to inefficacy or intolerability or those who remain in the study after initiating rescue medication. Mehrotra, et al. \cite{Mehrotra(2017)} discussed that the commonly used MMRM with the embedded MAR assumption can deliver an exaggerated estimate of the aforementioned estimand of interest, in favor of the drug. This happens, in part, due to implicit imputation of an overly optimistic mean for dropouts in the treatment group. To remedy this, they proposed a formula-based two-step approach by treating the true endpoint distribution for treatment group as a mixture of distributions (one each for the completers and dropouts) rather than a single distribution. Their approach reduces the bias associated with the traditional MMRM while maintaining power. To increase the precision in estimating $\delta$, we apply their method to MMRM, $\textrm{cLDA}^1$ and $\textrm{cLDA}^2$ models in the simulation study.

\section{Results}
\subsection{JMM and random forest fit to ADNI data}
\begin{figure}[H]
	\centering
	\includegraphics[scale = 0.07]{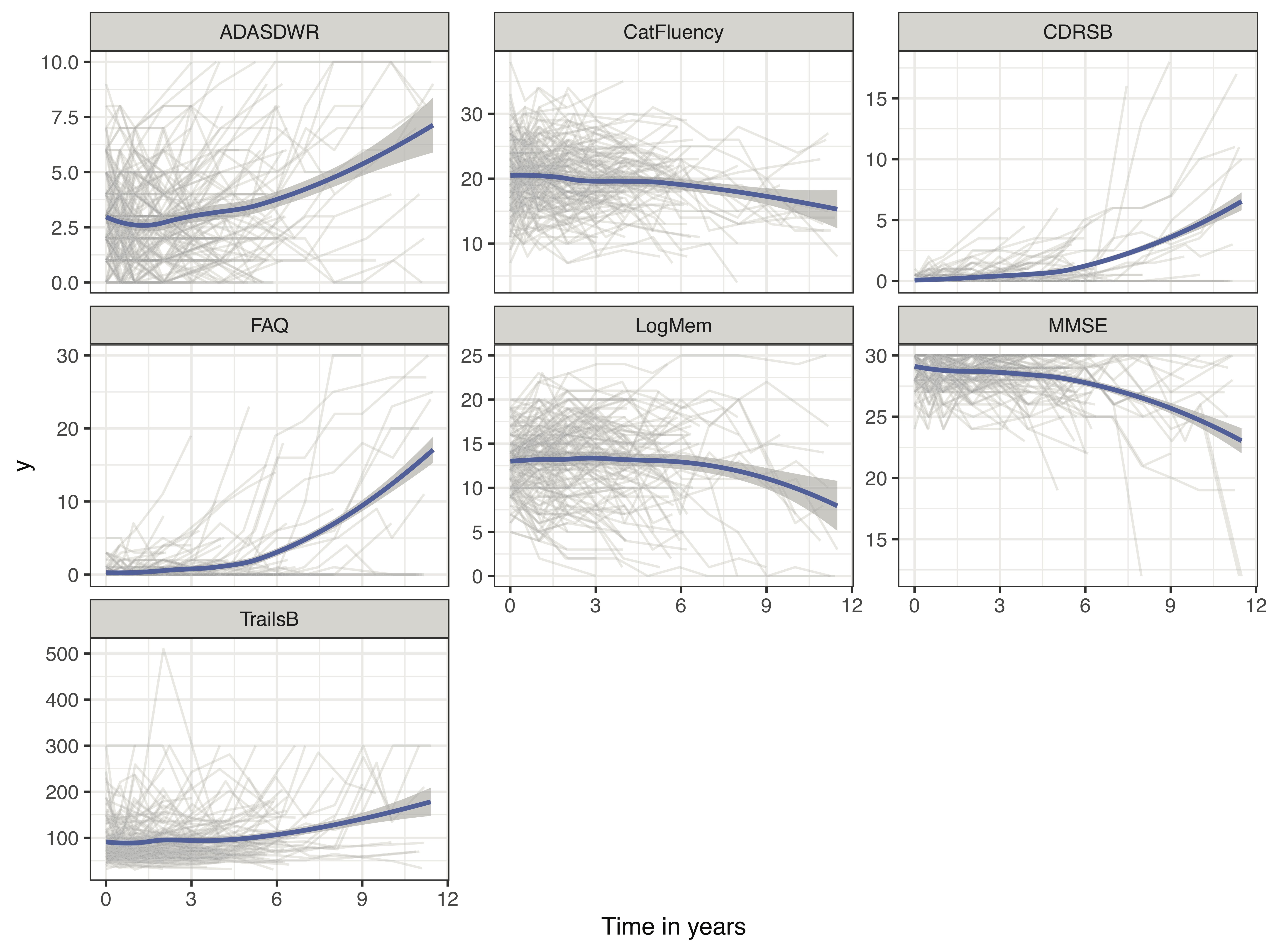}
	\includegraphics[scale = 0.07]{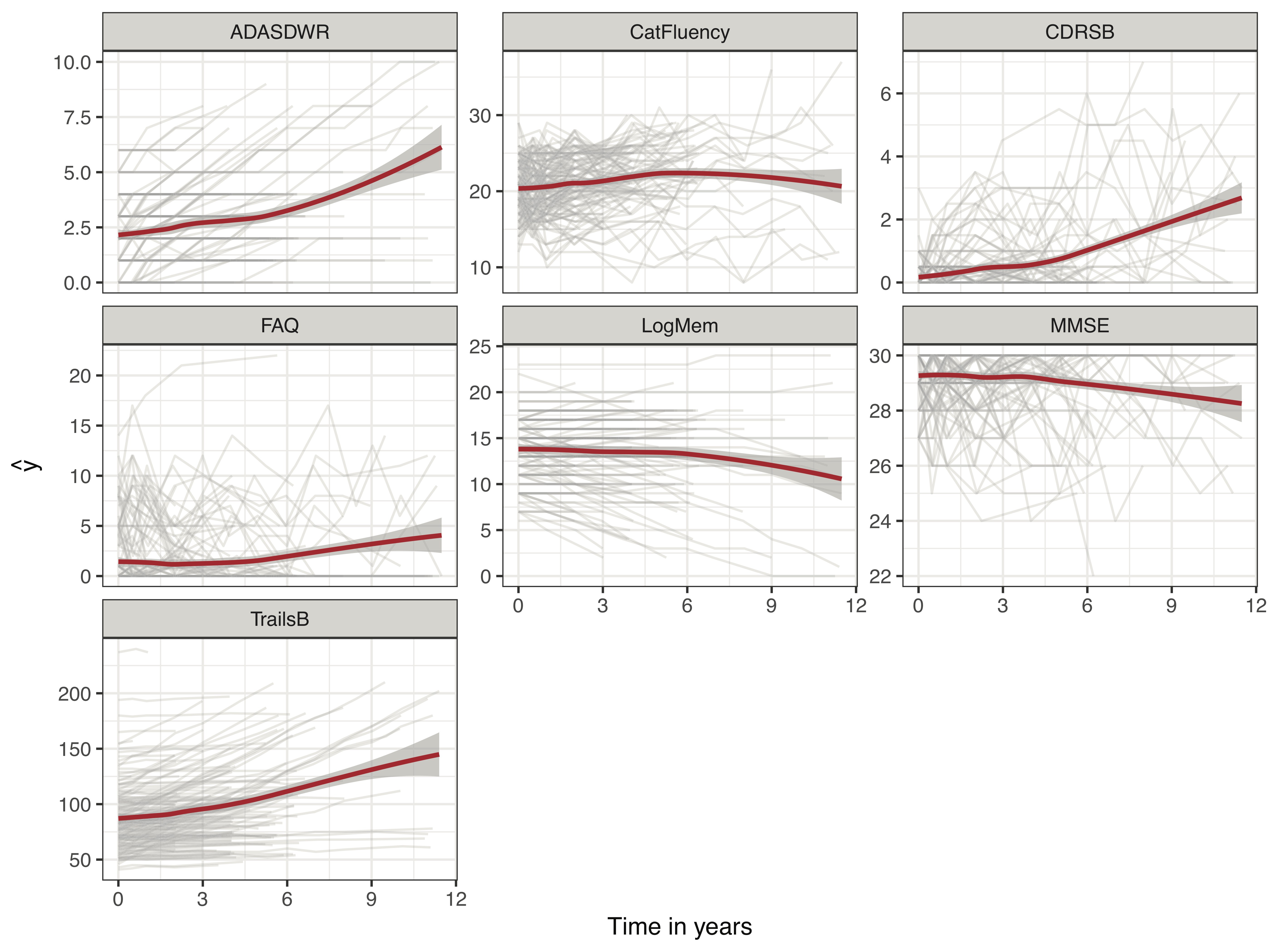}
	\caption{Observed (upper panel) and predicted (lower panel) longitudinal profiles of the seven markers for all individuals. Bold lines are LOESS smoothes.}
	\label{ADNI_observations}
\end{figure}

We fit a JMM for PAD participants who were observed to progress to MCI and a separate JMM for those who did not progress. Seven outcome measures described in Section \ref{data} are included in the model. Fixed effect covariates for each outcome include age at baseline and carriage of the \apoe allele. Three parallel Markov chains are run for 4000 iterations and the first 2000 warm-up iterations are discarded. Every fourth value of the remaining part of each chain is stored to reduce correlation, yielding a total of 1500 samples for posterior analysis. Table \ref{jm_estimates} shows the posterior means and 95\% credible intervals of the covariate-effect parameters. Figure \ref{ADNI_observations} shows the subject-level observations and predictions according to time in years of the seven markers for all individuals, in which the blue and red lines are the curves using the LOESS smoother. The bottom panel shows that the predictions provide reasonable trends of the observations. The posterior estimates from JMM will be later used as the true parameter values to simulate the panel of continuous markers.

\begin{figure}[H]
	\centering
	\includegraphics[scale=0.8]{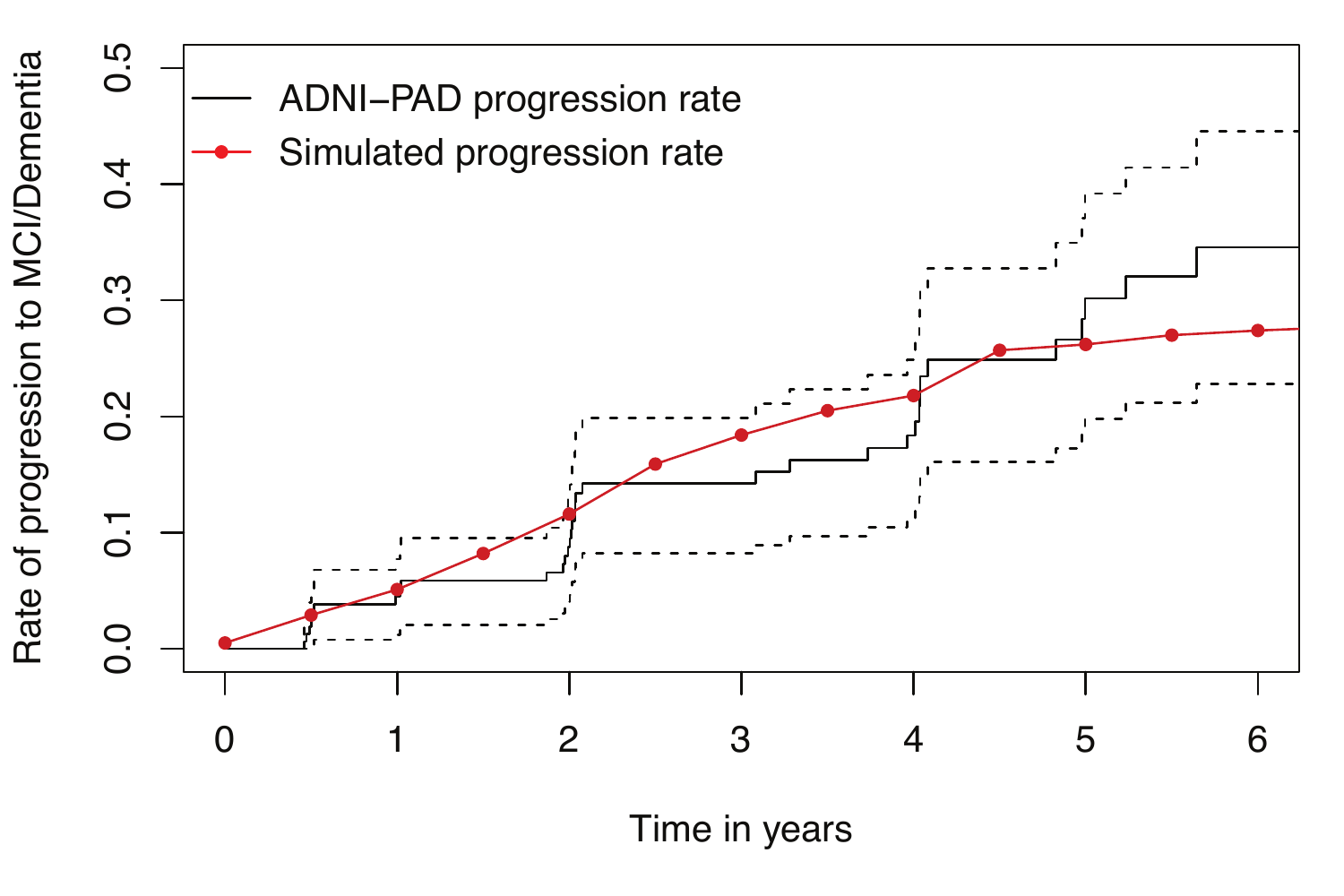}
	\caption{Kaplan-Meier estimated rate of progression to MCI or Dementia.}
	\label{Progression_rate}
\end{figure}

For the random forest, 500 trees are fitted and the number of variables selected at each split is 3. The node impurity of each tree is measured by the Gini index. The results show that CDRSB, LogMem and FAQ are three most important outcomes for determining the diagnosis of MCI. The model has a 6.19\% out-of-bag error rate and 93.81\% out-of-bag accuracy rate. Using the fitted random forest, the simulated cognitive status can be obtained from the simulated continuous markers. Figure \ref{Progression_rate} shows the Kaplan-Meier estimated progression rate of the ADNI-PAD population (black solid line) along with the progression rate from one large simulated placebo group (red dots). The simulated progression yields closer concordance with the Kaplan-Meier estimates at the earlier stage. Although we observe discrepancies between the two lines in the middle and the right tail, the red line still lies within the 95\% confidence intervals. Both the subject-level trajectories and the progression rate illustrate that the simulated data plausibly mimics the observed data.

%
%

\subsection{Simulation results}
\begin{figure}[H]
	\centering
	\hspace*{-0.75in}
	\includegraphics[scale = 0.45]{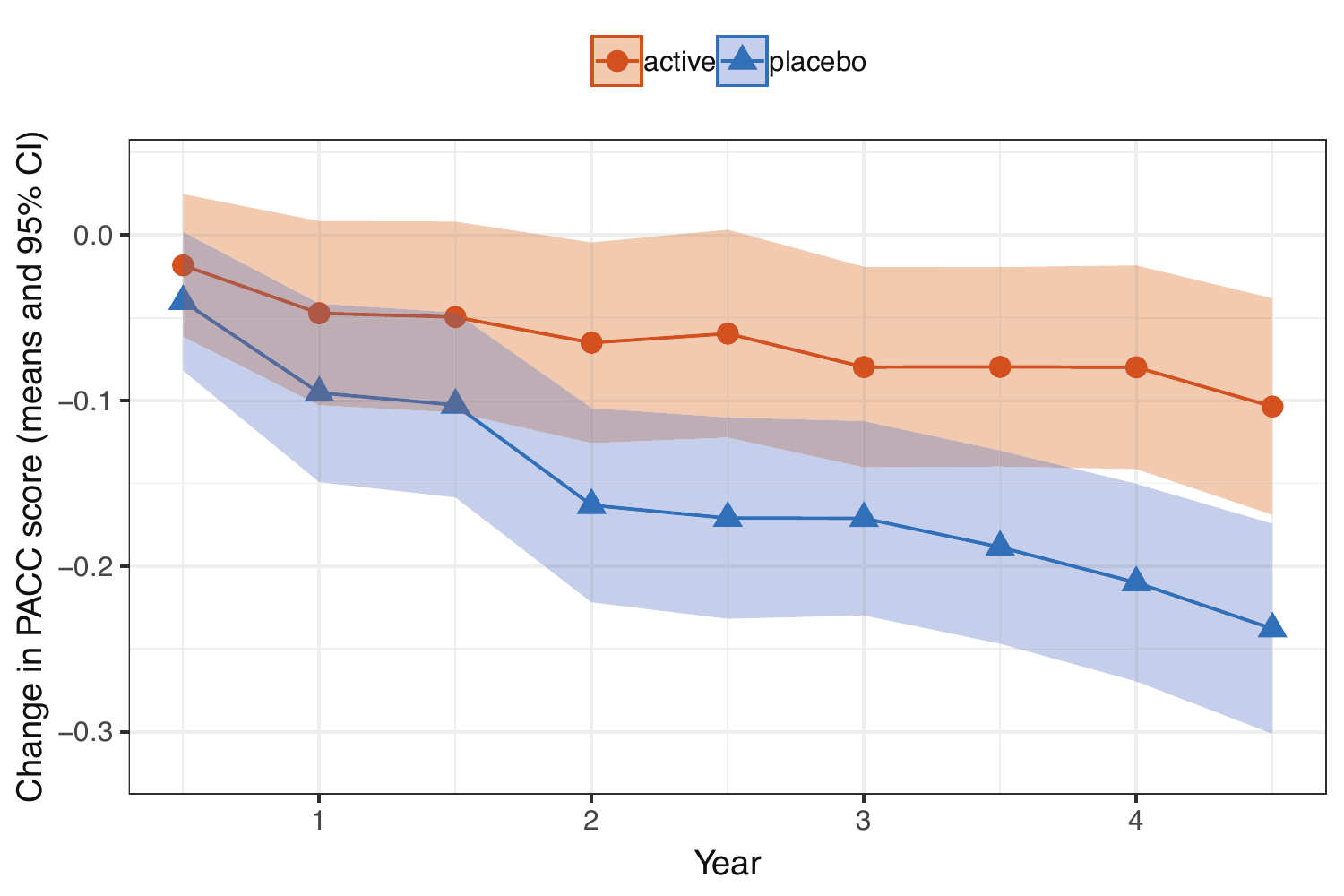}
	\includegraphics[scale = 0.45]{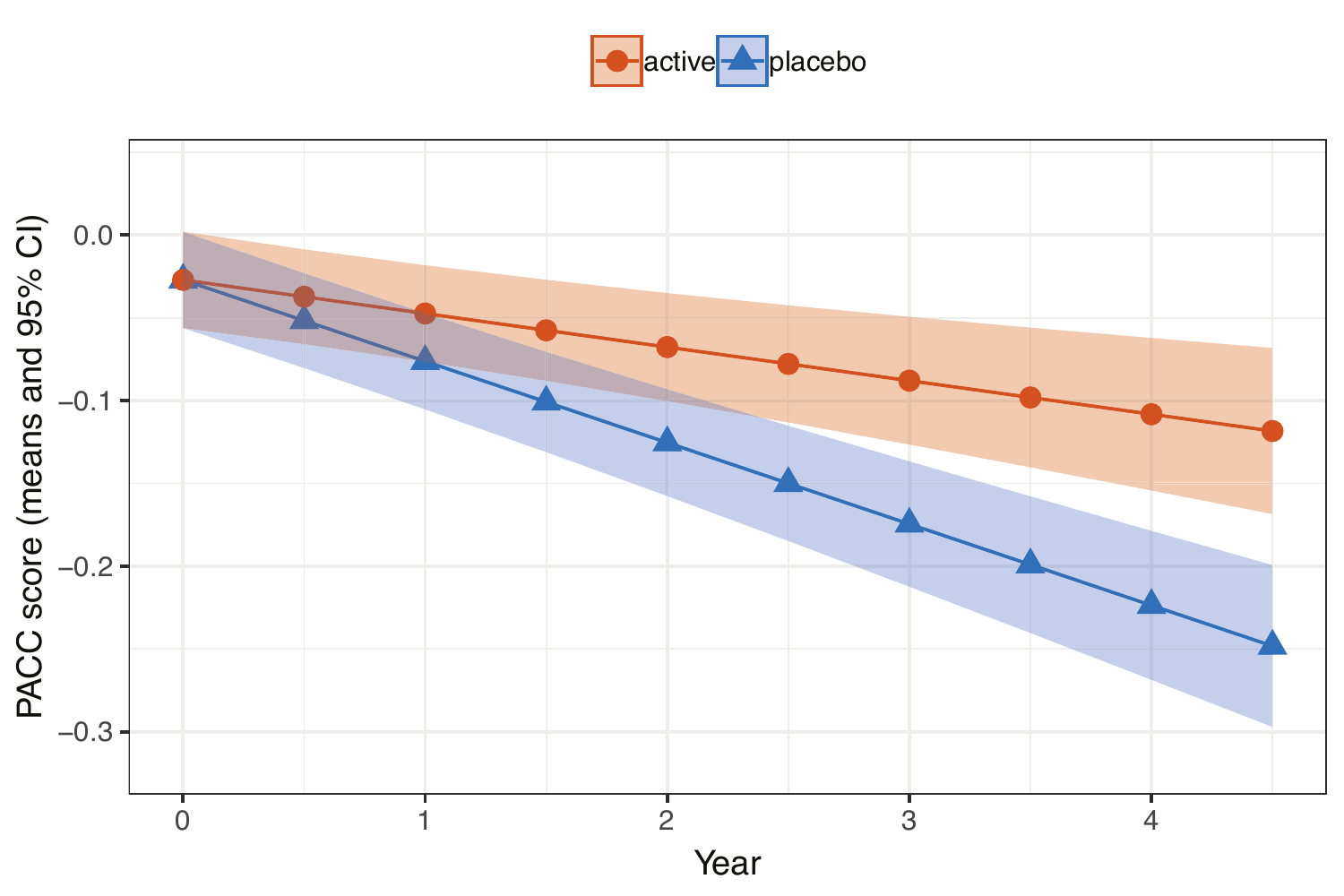}\\
	\hspace*{-0.75in}
	\includegraphics[scale = 0.45]{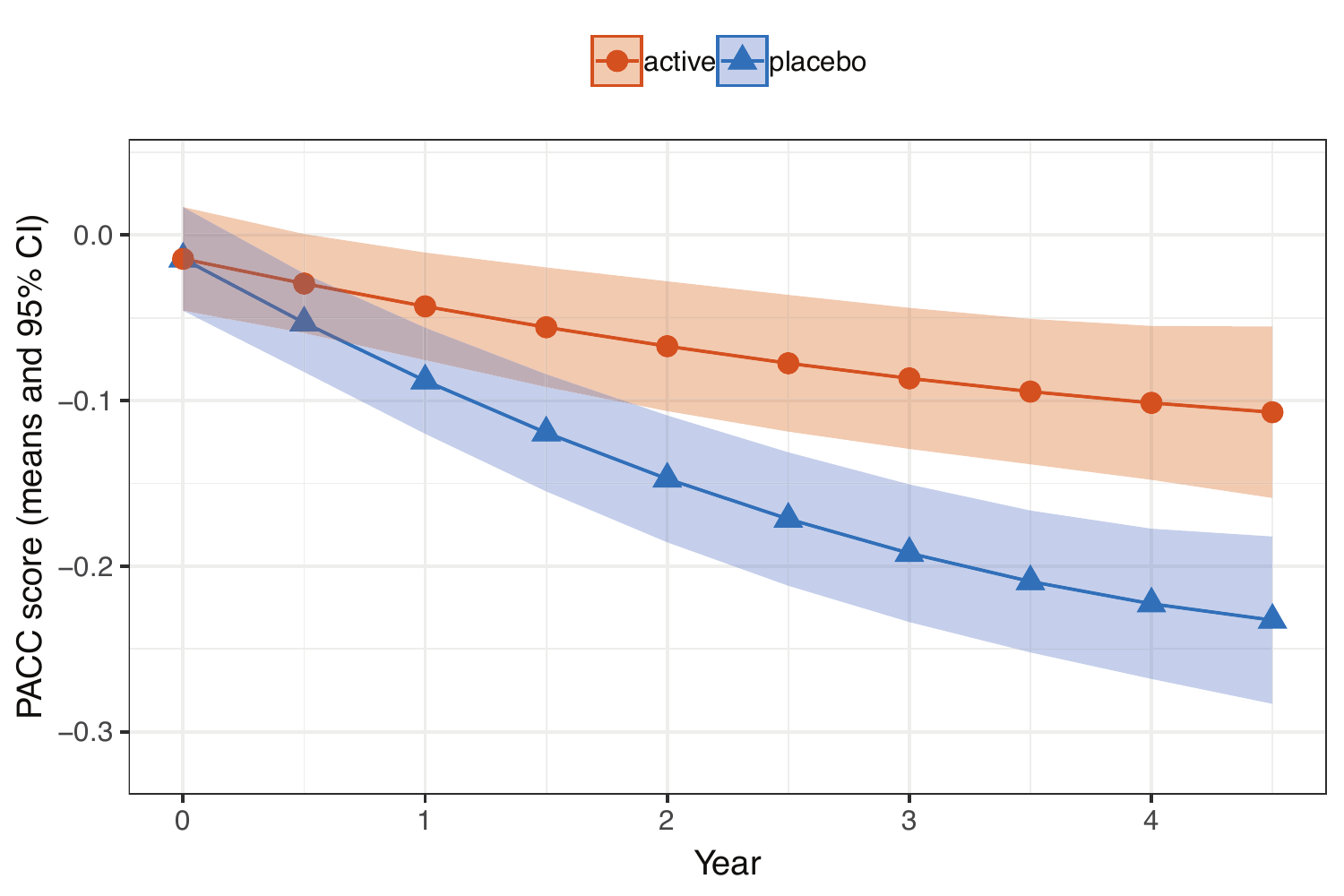}
	\includegraphics[scale = 0.45]{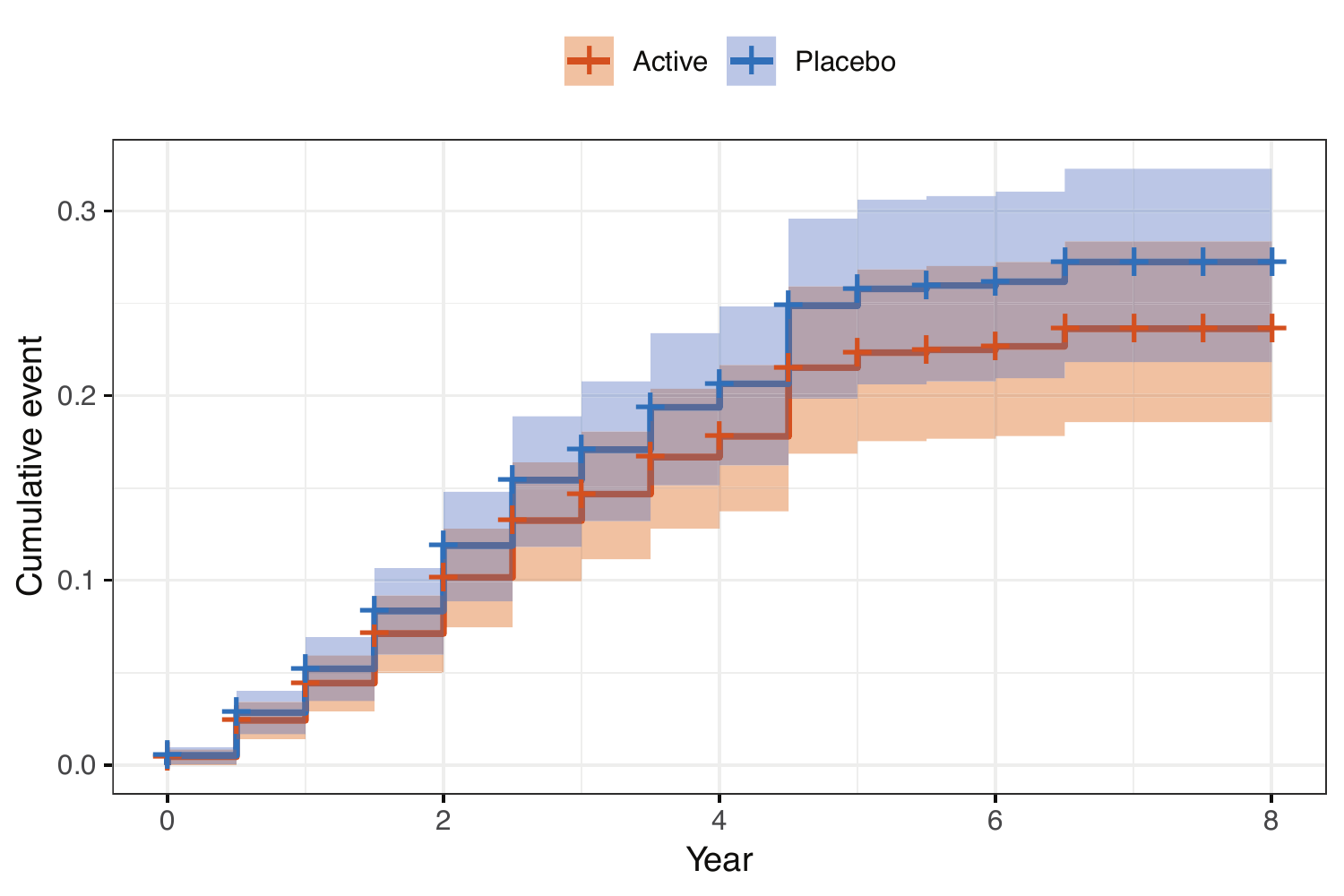}
	
	\caption{Results of one simulated clinical trial with 20\% treatment effect from (a) analysis of change from baseline using a categorical time MMRM of the PACC; (b) a cLDA model of PACC with linear time trends; (c) a cLDA model of PACC with quadratic time effects; and (d) Kaplan-Meier curves comparing the time-to-progression to Mild Cognitive Impairment or dementia for the two groups.}
	\label{plots_bygroup}
\end{figure}

Figure \ref{plots_bygroup} shows the results of one simulated clinical trial with a 20\% treatment effect and sample size $n=1000$. The figure illustrates the group trends obtained by fitting the four different models.

\begin{figure}[H]
	\centering
	\includegraphics[scale=0.55]{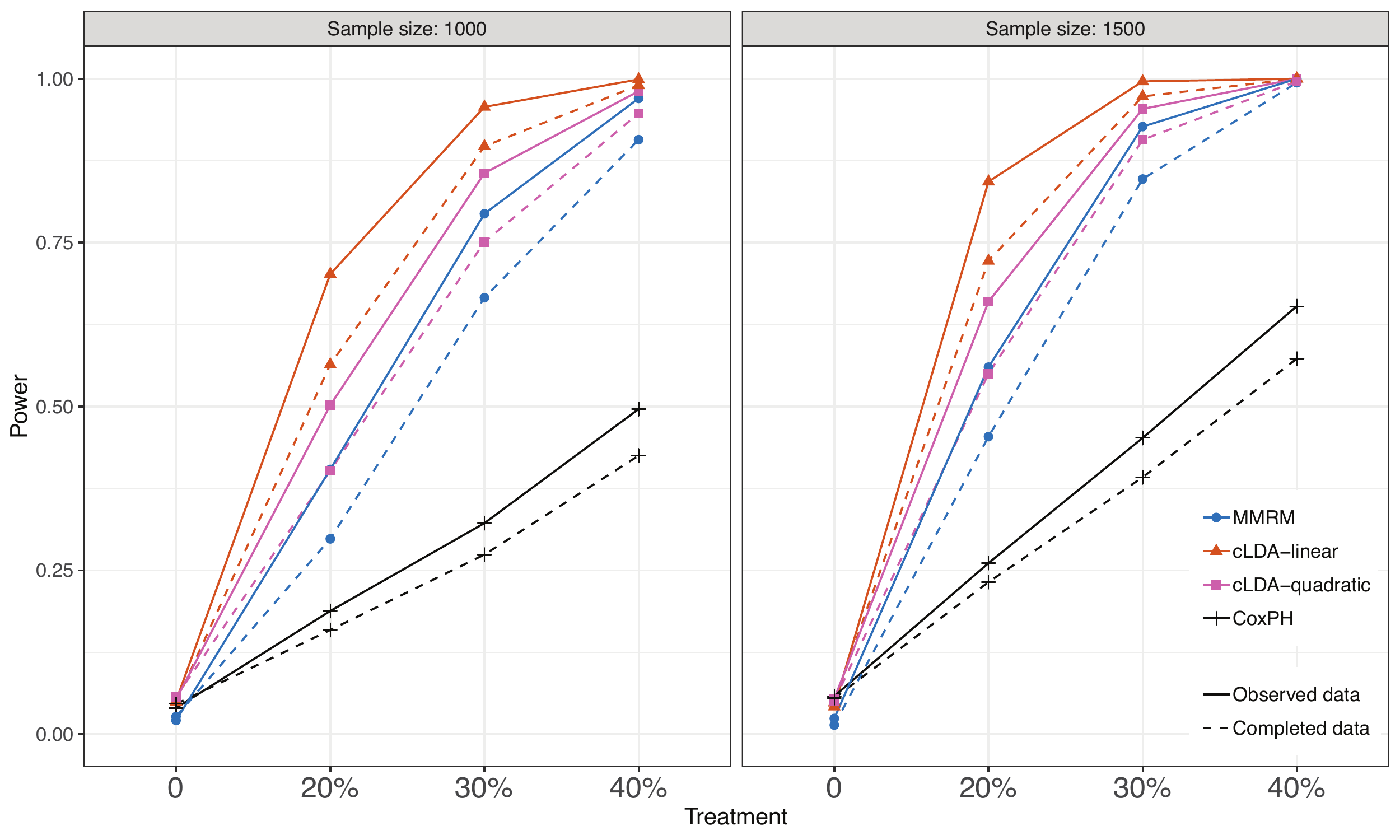}
	\caption{Statistical power for the MMRM, cLDA and Cox proportional hazards model for treatment effects 0\% (Type I error), 20\%, 30\% and 40\% for sample sizes of n=1000 (left panel) and n=1500 (right panel). Solid lines indicate power estimates for data observed after simulated non-ignorable missingness; and dashed lines indicate power that would be achieved with complete data (including observations that would be unobserved in reality). The observed data shows greater power with fewer observations because the non-ignorable missingness induces a bias in favor of the treatment.}
	\label{power}
\end{figure}

Simulated power and Type I error are summarized in Table \ref{powertab}. Under the null hypothesis (no treatment effect), the MMRM exhibits smaller than expected Type I error (about 2\%), whereas the other models are closer to the expect 5\% error rate. The Cox model consistently exhibits the weakest power of the four models. MMRM has the next best performance, followed by the quadratic (cLDA$^2$) and linear (cLDA$^1$) models. For example, with a trial of size $N=$1,000 subjects of drug with a 30\% treatment effect, the simulated power is 33\% for Cox, 79\% for MMRM, 86\% for cLDA$^2$, and 96\% for cLDA$^1$. In comparing analysis of complete versus observed data, it seems the missing data does not increase Type I error, but it does inflate power. This suggests the bias is only an issue with an effective drug, in which case the effectiveness might appear inflated. Figure \ref{power} shows the powers in all scenarios.

Tables \ref{biastab} and \ref{biaspercenttab} further examine the bias induced by the missing data pattern. The tables summarize the median and interquartile ranges $\left(\textrm{Q}_1,\textrm{Q}_3\right)$ of the bias on the PACC scale (Tables \ref{biastab}) and as a percent of effect seen in complete data (\ref{biaspercenttab}). The Cox model seems to have smaller bias with 20\% treatment effect, but as the treatment grows, the bias is comparable for all models. The method proposed by Mehrotra, et al. \cite{Mehrotra(2017)} successfully shrinks the magnitude of bias, e.g. from 27\% in favor of treatment to -4.4\% in favor of placebo for MMRM with 20\% treatment effect. The method appears to overcorrect the bias in favor of placebo in these simulations.

\section{Discussion}
\label{discussion}

We use Bayesian joint mixed effects models fit using ADNI data to simulate correlated longitudinal data that might plausibly arise in a PAD clinical trial. We used a random forest algorithm, also fit using ADNI, to algorithmically diagnose MCI in the simulated data so that we could compare models of the PACC to the Cox model of time-to-progression. The models of PACC consistently provide at least twice the power of the Cox model even when the Cox model has the benefit of considerably more follow-up under a common close design. Given this inefficiency, the time-to-progression analysis should be avoided in PAD.

Some might still argue that the clinical meaningfulness of the time-to-progression is worth the cost of a larger, longer trial. However, given that the random forest provided a purely algorithmic diagnosis with 93.81\% out-of-bag accuracy suggests that there is minimal additional value in the diagnosis. And again, while the progression outcome is more qualitative than the PACC on the subject level, the group level result is still quantitative (e.g. a hazard ratio) and requires additional interpretation to assign clinical meaning.

One might also argued that clinical diagnosis cannot be adequately modelled algorithmically using trial data. That is, clinical assessment and diagnosis by a trial site clinician may consider information not captured by trial measures. But the cognitive, clinical and functional assessments are designed to capture the relevant information, and clinicians generally rely on similar information obtained through less structured assessments. It seems questionable that a site clinician will gain much reliable information beyond the assessments; indeed, this is the justification for central expert panel adjudication of site diagnoses.

The Bayesian joint models are well-suited to simulating plausible panels of correlated longitudinal data necessary to compare clinical trial designs. This approach could be useful in many other contexts where one is interested in a fair comparison of different outcome measures, different combinations of correlated outcomes, or different models of treatment effect. Simulations which ignore the correlations among important outcomes will likely not provide reliable comparisons.

All of the models considered were susceptible to bias induced by a plausible missing data pattern. However, this bias seemed to only affect scenarios with an effective treatment and did not inflate Type I error under the null hypothesis. The Mehrotra method shows promise in correcting this bias, but it might overcorrect in favor of placebo, and it would be impossible to detect this overcorrection in practice. Given that Type I error is not inflated, we are inclined to suggest no change to the status quo approach in which the primary analysis is based on likelihood-based methods which are robust to MAR, and applying appropriate MNAR sensitivity analyses such as the delta method\cite{rubin1977formalizing}.

\section*{Conflicts of interest}
The authors declare no potential conflicts of interest.

\section*{Acknowledgments}
We are grateful to the ADNI study volunteers and their families. 

This work was supported by National Institute on Aging grant R01-AG049750. Data collection and sharing for this project was funded by the ADNI (National Institutes of Health Grant U01 AG024904) and DOD ADNI (Department of Defense award number W81XWH-12-2-0012). ADNI is funded by the National Institute on Aging, the National Institute of
Biomedical Imaging and Bioengineering, and through generous contributions from the following: AbbVie, Alzheimer's Association; Alzheimer's Drug Discovery Foundation; Araclon Biotech; BioClinica, Inc.; Biogen; Bristol-Myers Squibb Company; CereSpir, Inc.; Eisai Inc.; Elan Pharmaceuticals, Inc.; Eli Lilly and Company; EuroImmun; F. Hoffmann-La Roche Ltd and its affiliated company Genentech, Inc.; Fujirebio; GE Healthcare; IXICO Ltd.; Janssen Alzheimer Immunotherapy Research \& Development, LLC.; Johnson \& Johnson Pharmaceutical Research \& Development LLC.;
Lumosity; Lundbeck; Merck \& Co., Inc.; Meso Scale Diagnostics, LLC.; NeuroRx Research; Neurotrack Technologies; Novartis Pharmaceuticals Corporation; Pfizer Inc.; Piramal Imaging; Servier; Takeda Pharmaceutical Company; and Transition Therapeutics. The Canadian Institutes of Health Research provided funds to support ADNI clinical sites in Canada. Private sector contributions are facilitated by the Foundation for the National Institutes of Health (www.fnih.org). The grantee organization is the Northern California Institute for Research and Education, and the study is coordinated by the Alzheimer's Therapeutic Research Institute at the University of Southern California. ADNI data are disseminated by the Laboratory for Neuro Imaging at the University of Southern California.

\section*{Appendix}
For cLDA model with quadratic time effects, we can write the part of fixed effects as 
\begin{equation*}
y=\beta_0 + \beta_1 t + \beta_2 t \cdot \textrm{active} + \beta_3 t^2 + \beta_4 t^2 \cdot \textrm{active}
\end{equation*}
The area between the curves of active group and placebo group is 
\begin{equation*}
\begin{split}
S_{\textrm{trt-pb}} & = \int_{t_0}^{t_T}\left\{\beta_0+\left(\beta_1+\beta_2\right)t+\left(\beta_3+\beta_4\right)t^2 \right\} dt - \int_{t_0}^{t_T}\left\{\beta_0+\beta_1 t+\beta_3 t^2 \right\} dt\\
&= \left. \left(\frac{1}{2}\beta_2 t^2 + \frac{1}{3}\beta_4 t^3\right)\right|_{t=t_0}^{t=t_T}
\end{split}
\end{equation*}
The null hypothesis is $H_0$: $S_{trt-pb}=0$. We use the R package \texttt{glht} to carry out the hypothesis test.

\section*{References}

\bibliography{padbibfile}

\newpage 

\begin{table}[H]
	\centering
	\caption{Descriptive statistics by baseline diagnosis, normal cognition (NC) and subjective memory concern (SMC) for the Preclinical Alzheimer's Disease population in the Alzheimer's Disease Neuroimaging Initiative: Count (\%) or mean (SD).}
	\label{summaries}
	{
		\begin{tabular}{lrccc} \hline\hline		
			&  & \multicolumn{1}{c}{NC} & \multicolumn{1}{c}{SMC}  & \multicolumn{1}{c}{Total} \\
			Variable && \multicolumn{1}{c}{($N=120$)} & \multicolumn{1}{c}{($N=43$)} & \multicolumn{1}{c}{($N = 163$)} \\ \hline
			Age & & 75.21 (5.83) & 72.77 (5.78) & 74.57 (5.90)\\
			\apoe$\quad$alleles & 0 & 52 (43\%) & 23 (53\%) & 75 (46\%)\\
			& $\geq$1 & 111 (57\%) & 140 (47\%) & 88 (54\%)\\ 
			ADAS Delayed Word Recall & & 2.96 ( 1.79) & 3.00 ( 2.08) & 2.97 ( 1.86)\\
			Logical Memory - Delayed Recall & & 13.11 (  3.15) & 12.63 ( 3.19) & 12.98 ( 3.16)\\
			Trails B & & 93.40 (48.90) & 89.10 (32.00) & 92.30 (45.00)\\
			MMSE &  & 29.11 ( 1.13) & 29.09 ( 0.89) & 29.10 ( 1.07)\\
			Category Fluency (Animals) & & 20.72 ( 5.32) & 19.72 ( 5.60) & 20.45 ( 5.40)\\
			CDR-SB & 0 & 111 (92\%) &36 (84\%) & 147 (90\%) \\
			& 0.5 & 8 ( 7\%) & 7 (16\%) & 15 ( 9\%)\\
			& 1 & 1 ( 1\%) & 0 ( 0\%) & 1 ( 1\%)\\
			FAQ & 0 & 108 (90\%) &32 (74\%) & 140 (86\%)\\
			& 1 & 7 ( 6\%) & 8 (19\%) & 15 ( 9\%)\\
			& 2 & 2 ( 2\%) & 0 ( 0\%) & 2 ( 1\%)\\
			& 3 & 2 ( 2\%) & 3 ( 7\%) & 5 ( 3\%)\\
			& 5 & 1 ( 1\%) & 0 ( 0\%) & 1 ( 1\%)\\ \hline
		\end{tabular}
	}
\end{table}

\begin{table}[H]
	\centering
	\caption{Posterior estimates (means and 95\% credible intervals) of the fixed effect covariates for the joint mixed effect model fit to seven outcomes for stable and MCI progressor subpopulations.}
	\label{jm_estimates}
	{\small
		\begin{tabular}{rrrrrrr} \hline \hline
		     & \multicolumn{2}{c}{Progressor $\left(N=39\right)$} &&& \multicolumn{2}{c}{Stable $\left(N=124\right)$} \\ \cline{2-3}\cline{6-7}
		     Parameter & Mean & 95\% CI &&& Mean & 95\% CI \\ \hline
		     &&&&&&\\
		     \multicolumn{7}{l}{\underline{ADAS Delayed Word Recall}}\\
		     Intercept & -8.244 & $\left(-15.39,-1.451\right)$ &&& -4.913 & $\left(-7.755,-2.003\right)$ \\
		     Year & 0.330 & $\left(0.189,0.464\right)$ &&& 0.064 & $\left(0.021,0.108\right)$ \\
		     Age & 0.110 & $\left(0.021,0.201\right)$ &&& 0.062 & $\left(0.023,0.100\right)$ \\
		     \apoe & 0.572 & $\left(-0.319,1.437\right)$ &&& 0.218 & $\left(-0.247,0.670\right)$ \\
			 &&&&&&\\
		     \multicolumn{7}{l}{\underline{Logical Memory Paragraph Recall}}\\
		     Intercept & -6.897 & $\left(-15.425,0.905\right)$ &&& -1.840 & $\left(-4.983,1.350\right)$ \\
		     Year & 0.261 & $\left(0.136,0.395\right)$ &&& 0.033 & $\left(-0.084,0.016\right)$ \\
		     Age & 0.096 & $\left(-0.005,0.206\right)$ &&& 0.020 & $\left(-0.023,0.062\right)$ \\
		     \apoe & 0.039 & $\left(-0.959,1.099\right)$ &&& 0.465 & $\left(-0.044,0.985\right)$ \\
		     &&&&&&\\
		     \multicolumn{7}{l}{\underline{Trails B}}\\
		     Intercept & -9.458 & $\left(-14.898,-3.918\right)$ &&& -6.364 & $\left(-9.020,-3.792\right)$ \\
		     Year & 0.353 & $\left(0.252,0.445\right)$ &&& 0.022 & $\left(-0.028,0.073\right)$ \\
		     Age &0.124 & $\left(0.051,0.193\right)$ &&& 0.084 & $\left(0.050,0.119\right)$ \\
		     \apoe & 0.141& $\left(-0.540,0.858\right)$ &&& 0.622 & $\left(0.187,1.087\right)$ \\
		     &&&&&&\\
		     \multicolumn{7}{l}{\underline{Mini-Mental State Examination}}\\
		     Intercept & 0.852 & $\left(-191.780,185.973\right)$ &&& -1.385 & $\left(-75.020,72.568\right)$ \\
		     Year & 0.009 & $\left(-3.918,4.011\right)$ &&& 0.022 & $\left(-2.590,2.698\right)$ \\
		     Age & 0.007 & $\left(-2.432,2.436\right)$ &&& 0.020 & $\left(-0.903,0.944\right)$ \\
		     \apoe & 0.040 & $\left(-1.116,11.346\right)$ &&& 0.115 & $\left(-5.683,5.900\right)$ \\
		     &&&&&&\\
		     \multicolumn{7}{l}{\underline{Category Fluency - Animals}}\\
		     Intercept & 1.430& $\left(-127.590,130.195\right)$ &&& 0.942 & $\left(-96.958,98.426\right)$ \\
		     Year & 0.047& $\left(-2.910,2.786\right)$ &&& 0.025 & $\left(-3.399,3.798\right)$ \\
		     Age & -0.009 & $\left(-1.658,1.606\right)$ &&& -0.011 & $\left(-1.224,1.211\right)$ \\
		     \apoe & 0.036 & $\left(-8.234,8.775\right)$ &&& -0.118 & $\left(-7.911,7.920\right)$ \\
		     &&&&&&\\
		     \multicolumn{7}{l}{\underline{Clinical Dementia Rating - Sum of Boxes}}\\
		     Intercept & -6.537 & $\left(-364.967,344.177\right)$ &&& 1.094 & $\left(-82.421,76.732\right)$ \\
		     Year & 0.082 & $\left(-7.263,6.390\right)$ &&& 0.006 & $\left(-2.853,2.947\right)$ \\
		     Age & 0.081 & $\left(-4.230,4.517\right)$ &&& -0.011 & $\left(-1.006,1.027\right)$ \\
		     \apoe & -0.224 & $\left(-20.697,19.566\right)$ &&& 0.117 & $\left(-5.925,6.358\right)$ \\
		     &&&&&&\\
		     \multicolumn{7}{l}{\underline{Functional Assessment Questionnaire}}\\
		     Intercept & 3.458 & $\left(-380.068,367.151\right)$ &&& 0.261 & $\left(-32.960,32.991\right)$ \\
		     Year & 0.023 & $\left(-7.838,7.140\right)$ &&& 0.0007 & $\left(-1.1420,1.1710\right)$ \\
		     Age & -0.002 & $\left(-4.487,4.718\right)$ &&& -0.003 & $\left(-0.410,0.449\right)$ \\
		     \apoe & 0.343 & $\left(-22.127,22.506\right)$ &&& 0.014 & $\left(-2.667,2.525\right)$ \\\hline
		\end{tabular}
	}
\end{table}

\begin{table}[H]
	\centering
	\caption{Power and Type I error from 1000 simulated clinical trials. The rows with 0\% treatment effect simulate the Type I error, which we expect to be near 5\%.}
	\label{powertab}
	{\footnotesize
  \begin{tabular}{ccrrrrcrrrr} \hline\hline
  	Sample & &	\multicolumn{4}{c}{Observed data} && \multicolumn{4}{c}{Completed data} \\
  	\cline{3-6}\cline{8-11}
  	 size & Treatment & MMRM & $\textrm{cLDA}^1$ & $\textrm{cLDA}^2$ & CoxPH && MMRM & $\textrm{cLDA}^1$ & $\textrm{cLDA}^2$ & CoxPH \\ \hline
  	\multirow{4}{*}{1000} & 0\% & 0.021 & 0.051 & 0.053 & 0.040 && 0.027 & 0.049 & 0.057 & 0.046\\
  	& 20\% & 0.404 & 0.702 & 0.502 & 0.188 && 0.298 & 0.564 & 0.402 & 0.159\\
  	& 30\% & 0.794 & 0.957 & 0.856 & 0.322 && 0.666 & 0.897 & 0.751 & 0.274\\
  	& 40\%  & 0.970 & 0.999 & 0.981 & 0.496 && 0.907 & 0.990 & 0.947 & 0.425\\ \hline
  	\multirow{4}{*}{1500} & 0\% & 0.024 & 0.042 & 0.054 & 0.058 && 0.014 & 0.048 & 0.051 & 0.055\\
  	& 20\% & 0.560 & 0.843 & 0.660 & 0.261 && 0.454 & 0.722 & 0.550 & 0.232\\
  	& 30\% & 0.927 & 0.996 & 0.954 & 0.452 && 0.847 & 0.973 & 0.907 & 0.392\\
  	& 40\% & 1.000 & 1.000 & 1.000 & 0.653 && 0.994 & 1.000 & 0.996 & 0.573\\\hline	
  \end{tabular}
	}
\end{table}

\begin{table}[H]
	\centering
	\caption{Bias of the treatment effect due to missingness.}
	\label{biastab}
	{\scriptsize
		\begin{tabular}{clrrrrrrrr} \hline\hline
			Sample & & \multicolumn{2}{c}{20\%} && \multicolumn{2}{c}{30\%} && \multicolumn{2}{c}{40\%} \\ \cline{3-4}\cline{6-7}\cline{9-10}
			size & Analysis Method & Median & $\left(\text{Q}_1, \text{Q}_3\right)$ && Median & $\left(\text{Q}_1, \text{Q}_3\right)$ && Median & $\left(\text{Q}_1, \text{Q}_3\right)$\\ \hline
			\multirow{8}{*}{1000} & MMRM & 0.018 & $\left(0.006,0.031\right)$ && 0.028 & $\left(0.015,0.040\right)$ && 0.037 & $\left(0.024,0.049\right)$\\
			& $\textrm{cLDA}^1$ & 0.019 & $\left(0.009,0.029\right)$ && 0.028 & $\left(0.018,0.038\right)$ && 0.038 & $\left(0.028,0.048\right)$\\
			& $\textrm{cLDA}^2$ & 0.038 & $\left(0.011,0.065\right)$ && 0.058 & $\left(0.030,0.084\right)$ && 0.077 & $\left(0.050,0.104\right)$\\
			& CoxPH & -0.033 & $\left(-0.074,0.010\right)$ && -0.045 & $\left(-0.086,-0.001\right)$ && -0.059 & $\left(-0.102,-0.017\right)$\\
			& MMRM-Mehrotra & -0.001 & $\left(-0.015,0.012\right)$ && -0.002 & $\left(-0.016,0.011\right)$ && -0.003 & $\left(-0.017,0.010\right)$\\
			& $\textrm{cLDA}^1$-Mehrotra & -0.001 & $\left(-0.011,0.008\right)$ && -0.001 & $\left(-0.012,0.007\right)$ && -0.003 & $\left(-0.012,0.007\right)$\\
			& $\textrm{cLDA}^2$-Mehrotra & -0.006 & $\left(-0.034,0.022\right)$ && -0.010 & $\left(-0.038,0.018\right)$ && -0.014 & $\left(-0.042,0.014\right)$\\\hline
			\multirow{8}{*}{1500} & MMRM & 0.018 & $\left(0.006,0.028\right)$ && 0.027 & $\left(0.016,0.037\right)$ && 0.036 & $\left(0.025,0.047\right)$\\
			& $\textrm{cLDA}^1$ & 0.018 & $\left(0.010,0.026\right)$ && 0.027 & $\left(0.019,0.036\right)$ && 0.037 & $\left(0.028,0.045\right)$\\
			& $\textrm{cLDA}^2$ & 0.037 & $\left(0.013,0.061\right)$ && 0.056 & $\left(0.032,0.080\right)$ && 0.075 & $\left(0.052,0.099\right)$\\
			& CoxPH & -0.028 & $\left(-0.064,0.005\right)$ && -0.042 & $\left(-0.076,-0.009\right)$ && -0.055 & $\left(-0.090,-0.021\right)$\\
			& MMRM-Mehrotra & -0.002 & $\left(-0.012,0.009\right)$ && -0.003 & $\left(-0.013,0.008\right)$ && -0.004 & $\left(-0.015,0.007\right)$\\
			& $\textrm{cLDA}^1$-Mehrotra & -0.001 & $\left(-0.009,0.006\right)$ && -0.002 & $\left(-0.010,0.005\right)$ && -0.003 & $\left(-0.011,0.004\right)$\\
			& $\textrm{cLDA}^2$-Mehrotra & -0.008 & $\left(-0.028,0.015\right)$ && -0.012 & $\left(-0.032,0.011\right)$ && -0.016 & $\left(-0.035,0.007\right)$\\\hline
		\end{tabular}
	}
\end{table}

\begin{table}[H]
	\centering
	\caption{Bias in percent (\%) of the treatment effect due to missingness based on 1000 simulated trials for the given sample size, treatment effect, and analysis method.}
	\label{biaspercenttab}
	{\footnotesize
		\begin{tabular}{clrrrrrrrr} \hline\hline
			Sample & & \multicolumn{2}{c}{20\%} && \multicolumn{2}{c}{30\%} && \multicolumn{2}{c}{40\%} \\ \cline{3-4}\cline{6-7}\cline{9-10}
			size & Analysis Method & Median & $\left(\text{Q}_1, \text{Q}_3\right)$ && Median & $\left(\text{Q}_1, \text{Q}_3\right)$ && Median & $\left(\text{Q}_1, \text{Q}_3\right)$\\ \hline
			\multirow{8}{*}{1000} & MMRM & 27.1 & $\left(7.0,52.3\right)$ && 29.9 & $\left(16.3,46.8\right)$ && 29.6 & $\left(19.4,42.3\right)$\\
			& $\textrm{cLDA}^1$ & 29.6 & $\left(12.4,51.9\right)$ && 29.8 & $\left(18.9,43.7\right)$ && 29.7 & $\left(21.4,39.7\right)$\\
			& $\textrm{cLDA}^2$ & 24.5 & $\left(5.5,50.2\right)$ && 26.5 & $\left(13.7,42.6\right)$ && 26.2 & $\left(16.5,37.9\right)$\\
			& CoxPH & 17.4 & $\left(-16.1,55.0\right)$ && 22.2 & $\left(-4.5,52.7\right)$ && 25.5 & $\left(5.2,50.4\right)$\\
			& MMRM-Mehrotra & -4.4 & $\left(-23.2,20.6\right)$ && -2.9 & $\left(-15.9,13.3\right)$ && -2.8 & $\left(-12.7,8.6\right)$\\
			& $\textrm{cLDA}^1$-Mehrotra & -1.7 & $\left(-16.2,15.4\right)$ && -1.7 & $\left(-11.3,9.1\right)$ && -2.0 & $\left(-9.2,5.7\right)$\\
			& $\textrm{cLDA}^2$-Mehrotra & -6.0 & $\left(-21.2,15.9\right)$ && -4.5 & $\left(-15.5,9.4\right)$ && -4.7 & $\left(-13.0,5.2\right)$\\\hline
			\multirow{8}{*}{1500} & MMRM & 27.5 & $\left(9.7,52.8\right)$ && 28.2 & $\left(16.6,43.3\right)$ && 28.3 & $\left(19.6,39.3\right)$\\
			& $\textrm{cLDA}^1$ & 29.1 & $\left(15.7,48.4\right)$ && 29.2 & $\left(19.9,40.9\right)$ && 29.3 & $\left(22.2,37.8\right)$\\
			& $\textrm{cLDA}^2$ & 24.8 & $\left(8.8,45.6\right)$ && 25.4 & $\left(15.2,38.2\right)$ && 25.5 & $\left(17.8,34.6\right)$\\
			& CoxPH & 18.0 & $\left(-8.2,46.9\right)$ && 22.7 & $\left(3,46.3\right)$ && 24.3 & $\left(8.6,44.6\right)$\\
			& MMRM-Mehrotra & -3.0 & $\left(-19.4,17.6\right)$ && -3.0 & $\left(-13.8,9.7\right)$ && -3.1 & $\left(-11.2,6.2\right)$\\
			& $\textrm{cLDA}^1$-Mehrotra & -2.1 & $\left(-13.5,11.4\right)$ && -2.3 & $\left(-9.8,5.7\right)$ && -2.4 & $\left(-7.9,3.5\right)$\\
			& $\textrm{cLDA}^2$-Mehrotra & -6.1 & $\left(-18.8,12.7\right)$ && -5.5 & $\left(-13.9,5.7\right)$ && -5.5 & $\left(-11.5,2.8\right)$\\\hline
		\end{tabular}
	}
\end{table}

\end{document}